\documentclass[twoside]{article}

\usepackage[accepted]{test}

 \usepackage[round]{natbib}
\renewcommand{\bibname}{References}
\renewcommand{\bibsection}{\subsubsection*{\bibname}}

\usepackage{hyperref}
\hypersetup{colorlinks=true,linkcolor=red,citecolor = blue, linktocpage}

\usepackage{booktabs}       
\usepackage{amsfonts}       
\usepackage{nicefrac}       
\usepackage{microtype}      
\usepackage{enumerate}
\usepackage{caption}
\usepackage{subcaption}
\usepackage{bm}
\usepackage{textcomp}
\usepackage{graphicx}
\usepackage{epstopdf}
\usepackage{amssymb}
\usepackage{amsthm}
\usepackage{algorithmic}
\usepackage{algorithm}
\usepackage{clrscode}
\allowdisplaybreaks
\hypersetup{final}

\usepackage{mathrsfs}
\usepackage{psfrag}
\usepackage{tikz}
\usepackage{url}            

\usepackage{xcolor}

\newcommand{\nn }{\nonumber}

\newcommand{\be}{\begin{equation}}
\newcommand{\ee}{\end{equation}}
\makeatletter
\newtheorem*{rep@theorem}{\rep@title}
\newcommand{\newreptheorem}[2]{%
\newenvironment{rep#1}[1]{%
 \def\rep@title{#2 \ref{##1}}%
 \begin{rep@theorem}}%
 {\end{rep@theorem}}}
\makeatother

\newtheorem{theorem}{Theorem}
\newreptheorem{theorem}{Theorem}

\newreptheorem{lemma}{Lemma}

\newtheorem{proposition}{Proposition}
\newreptheorem{proposition}{Proposition}

\newtheorem{corollary}{Corollary}

\newtheorem{remark}{Remark}

\begin{document}

%

%

\runningauthor{Gholamali Aminian$^*$, Mahed Abroshan$^*$}

\runningtitle{An Information-theoretical Approach to Semi-supervised Learning under Covariate-shift}

\twocolumn[

\aistatstitle{An Information-theoretical Approach to Semi-supervised Learning under Covariate-shift}


\aistatsauthor{ Gholamali Aminian$^{\star\dagger}$ \quad Mahed Abroshan$^{\star\ddagger}$}\aistatsauthor{ Mohammad Mahdi Khalili$^{\dagger\dagger}$ \quad Laura Toni$^{\dagger}$ \quad Miguel R. D. Rodrigues$^{\dagger}$}

\aistatsaddress{$^\dagger$ University College London \And  $^{\ddagger}$ Alan Turing Institute \And $^{\dagger\dagger}$ University of Delaware}
]

\begin{abstract}
A common assumption in semi-supervised learning is that the labeled, unlabeled, and test data are drawn from the same distribution. However, this assumption is not satisfied in many applications. In many scenarios, the data is collected sequentially (e.g., healthcare) and the distribution of the data may change over time often exhibiting so-called covariate shifts. In this paper, we propose an approach for semi-supervised learning algorithms that is capable of addressing this issue. Our framework also recovers some popular methods, including entropy minimization and pseudo-labeling. We provide new information-theoretical based generalization error upper bounds inspired by our novel framework. Our bounds are applicable to both general semi-supervised learning and the covariate-shift scenario. Finally, we show numerically that our method outperforms previous approaches proposed for semi-supervised learning under the covariate shift.
\end{abstract}

\section{INTRODUCTION}
There are many applications, from natural language processing to bio-informatics in which the labeled data is scarce, while plenty of unlabeled data is accessible. Under these circumstances, semi-supervised learning (SSL) algorithms allow us to take advantage of both labeled and unlabeled data. There are different approaches to SSL~\citep{yang2021survey}, e.g., self-training or input-consistency regularization. Self-training algorithms~\citep{ouali2020overview}, which are also the focus of this paper, are a class of SSL approaches. These algorithms predict the label of unlabeled data using the model’s own confident
predictions. However, most of self-training SSL algorithms are designed by assuming labelled training data, and test data admit the same distribution.

There are though many applications in which unlabeled features have a different distribution from labeled features, and the test features distribution could be different or the same as unlabeled features distribution. This situation -- commonly known as covariate-shift -- arises where data is collected sequentially. For example, in many healthcare applications, the labels will be available with a delay (e.g., studying five years survival analysis or drug discovery) and the distribution of new unlabeled data may change~\citep{ryan2015semi}. Another scenario associated with covariate shift relates to cases where we have a limited number of labeled data on a particular task, and there is plenty of unlabeled data on some other related tasks~\citep{oliver2018realistic}. Studying generalization error bounds is critical to understanding SSL models' performance -- and designing suitable SSL approaches -- in the presence of the aforementioned distribution shifts.

Various approaches have been developed to characterize the generalization error of SSL algorithms. SSL Generalization error upper bound using Bayes classifiers are provided in  \citep{gopfert2019can} and \citep{zhu2020semi}.
The VC-dimension approach is applied by \citep{gopfert2019can} for the SSL algorithm.
\citep{zhu2020semi} provides an upper bound on the excess risk of SSL algorithm by considering an exponentially concave function\footnote{A function $f(x)$ is called $\beta$-exponentially concave function if $\exp(-\beta f(x))$ is concave} based on conditional mutual information. The generalization error of iterative SSL algorithms based on pseudo-labels is studied by \cite{he2021information}. Generalization error upper bound based on Rademacher complexity for binary classification with a squared-loss mutual information regularization is provided in \citep{niu2013squared}. An upper bound on the generalization error of binary classification under cluster assumption is derived in \citep{rigollet2007generalization}. We refer to the survey paper~\citep{mey2019improvability} and references therein for a thorough review of other theoretical aspects of SSL.

However, these upper bounds on excess risk and generalization error do not entirely capture the role of covariate-shift between labeled data and unlabeled data thereby limiting our ability to characterize the performance of existing SSL methods or to design new ones. In this paper, an information-theoretic approach, inspired by \cite{xu2017information} and \cite{russo2019much}, is applied to characterize the generalization ability of various self-training based SSL algorithms. Such approaches often express the generalization error in terms of certain information measures between the learning algorithm input (the training dataset including labeled and unlabeled data) and output (the hypothesis), thereby incorporating the various ingredients associated with the SSL problem, including the labeled and unlabeled data distribution, the hypothesis space, and the learning algorithm itself. Finally, inspired by our framework and the upper bounds on generalization error, we propose a new SSL algorithm that is able to take advantage of unlabeled data under covariate-shift.

Our main contributions are as follows:
\begin{itemize}
    \item We propose a novel framework for self-training SSL algorithms that encompasses traditional SSL approaches such as the entropy minimization and the Pseudo-labeling approaches. Our framework is applicable to different loss functions beyond the typically used log-loss function.

    \item We provide an information-theoretical upper bound on the expected generalization error of the SSL algorithms under covariate-shift in terms of KL divergence and total variation distance. We show that the unlabeled data in our framework can improve the generalization error convergence rate. 
    
    \item We provide novel information-theoretical upper bounds on \emph{the estimation error of true conditional probabilities} in terms of KL divergence and total variation distance, essential in self-training approaches.
    
    \item Inspired by our theoretical results, we then propose a method for SSL algorithm which outperforms traditional SSL algorithms in the presence of covariate shifts.

\end{itemize}

\textbf{Notations:} We adopt the following notation in the sequel. Upper-case letters denote random variables (e.g., $Z$), lower-case letters denote random variable realizations (e.g. $z$), and calligraphic letters denote spaces (e.g. $\mathcal{Z}$). We denote the distribution of the random variable $Z$ by $P_Z$, the joint distribution of two random variables $(Z_1,Z_2)$ by $P_{Z_1,Z_2}$.

\textbf{Information Measures:} If $P$ and $Q$ are probability measures over space $\mathcal{Z}$, and $P$ is absolutely continuous with respect to $Q$, the Kullback-Leibler (KL) divergence between $P$ and $Q$ is given by
$D(P\|Q)\triangleq\int_\mathcal{Z}\log\left(\frac{dP}{dQ}\right) dP$. If $Q$ is also absolutely continuous with respect to $P$, then the KL divergence is bounded. The entropy of probability measure, $P$, is given by $H(P)=\int_\mathcal{Z} -dP \log(dP)$.

The mutual information between two random variables $Z$ and $T$ is defined as the KL divergence between the joint distribution and product-of-marginal
distributions $I(Z;T)\triangleq D(P_{Z,T}\|P_Z\otimes P_{T})$, or equivalently, the conditional KL divergence between $P_{T|Z}$ and $P_T$ averaged over $P_Z$, $D(P_{T|Z} \| P_T|P_{Z})\triangleq\int_\mathcal{Z}D(P_{T|Z=z} \| P_T) dP_{Z}(z)$.

The total variation distance for two probability measures, $P$ and $Q$, is defined as
\begin{equation}
    \mathbb{TV}(P,Q)=\frac{1}{2}\int_{\mathcal{Z}} |dP-dQ|
\end{equation}
and the variational representation of total variation distance is as follows~\citep{polyanskiy2014lecture}:
\begin{equation}\label{Eq: tv rep}
    \mathbb{TV}(P,Q)=\frac{1}{2L}\sup_{g \in \mathcal{G}_L}\left\{\mathbb{E}[g(P)]-\mathbb{E}[g(Q)]\right\}
\end{equation}
where $\mathcal{G}_L=\{g: \mathcal{Z}\rightarrow \mathbb{R}, ||g||_\infty \leq L \}$. Note that the total variation is bounded, $\mathbb{TV}(P,Q) \leq 1$.

\section{RELATED WORK}
We now highlight some of the key relevant works in the surrounding fields, including SSL, covariate-shift, domain adaptation, and information-theoretic based generalization error upper bounds.

\textbf{Semi-Supervised Learning:} Entropy minimization and Pseudo-labeling are two fundamental approaches in self-training based SSL. 
In entropy minimization, an entropy function of the predicted conditional distribution is added to the main empirical risk function, which depends on unlabeled data~ \citep{grandvalet2005semi}. The entropy function can be viewed as a regularization term that penalizes uncertainty in the prediction of the label of the unlabelled data. There are some assumptions for the performance of entropy minimization algorithm, including manifold assumption~\citep{iscen2019label}-- where it is assumed that labelled and unlabelled features are drawn from a common data manifold -- or cluster assumptions~\citep{chapelle2003cluster}-- where it is assumed that similar data features have a similar label.
In contrast, in Pseudo-labeling, the model is trained using labeled data in a supervised manner and then used to provide a pseudo-label for the unlabeled data with high confidence~\citep{lee2013pseudo}. These pseudo labels are then used as inputs in another model, which is trained based on labeled and pseudo-labeled data in a supervised manner. However, pseudo-labelling approaches can underperform because they largely rely on the accuracy of the pseudo-labeling process. To bypass this challenge, an uncertainty-aware Pseudo-labeling approach is proposed in \citep{rizve2020defense}. A theoretical framework for using input-consistency regularization combined with self-training algorithms in deep neural networks is proposed in \citep{wei2020theoretical}. Some works also discuss how to combine SSL with causal learning~\citep{scholkopf2012causal,janzing2010causal,janzing2015semi}. 

Our work departs from existing SSL literature a novel framework -- which encompasses existing SSL methods such as entropy minimization or pseudo-labelling -- that can extended to other loss function. We also propose to consider the labeled data in the unsupervised loss function in order to improve SSL performance further.

\textbf{Covariate-shift:} Covariate-Shift has been studied in supervised learning~\citep{sugiyama2007covariate} and~\citep{shimodaira2000improving} and SSL scenarios~\citep{kawakita2013semi}. An approximate Bayesian inference scheme by using posterior regularisation for SSL in the presence of covariate-shift is provided in \citep{chan2020unlabelled}. The performance of self-training for the SSL algorithms, including entropy minimization and pseudo-labeling in the presence of covariate-shift, with spurious features, is studied in \citep{chen2020self}. Our work differs from this body of research in the sense that we provide an algorithm dependent upper bound on the generalization error of SSL algorithms under covariate-shift.

\textbf{Domain Adaptation:} 
Domain adaptation involves training a model based on labeled data from the source domain and unlabeled data from the target domain. The covariate-shift reduces to the domain adaptation scenario by considering the same conditional distribution of label given feature but different marginal distributions for the features and unlabeled data. The work \citep{ben2010theory}, proposed $\mathcal{H}\Delta \mathcal{H}$-divergence as a similarity metric, and generalization bound based on VC-dimension approach is provided. The authors in \citep{mansour2009domain} proposed the discrepancy distance for the general loss function, and a generalization bound based on the Rademacher complexity is derived. Inspired by \cite{ben2010theory}, a domain adversarial algorithm which minimizes $\mathcal{H}\Delta \mathcal{H}$-divergence between source and target domains, is provided by \cite{ganin2016domain}. The application of entropy minimization and the combination of domain adversarial and entropy minimization in domain adaptation are proposed in \citep{wang2020tent} and \citep{shu2018dirt}. Our work differs from this area of research as our theoretical results are derived based on the covariate-shift assumption, i.e., the same conditional distribution of features given data under labeled and unlabeled data. In addition,  the estimation of conditional distributions in the domain adversarial  approach is induced by mostly labeled data. However, in our algorithm, this estimation is induced by both labeled and unlabeled data. 


\textbf{Information-theoretic upper bounds:} 
Recently, \cite{russo2019much,xu2017information} proposed to use the mutual information between the input training set and the output hypothesis to upper bound the expected generalization error. \cite{bu2020tightening} provides tighter bounds by considering the individual sample mutual information, \citep{asadi2018chaining} proposes using chaining mutual information, and some works advocate the conditioning and processing techniques~\citep{steinke2020reasoning,hafez2020conditioning,haghifam2020sharpened}. 
Information-theoretic generalization error bounds using other information quantities are also studied, such as $\alpha$-R\'enyi divergence and maximal leakage~\citep{esposito2019generalization}, Jensen-Shannon divergence~\citep{aminian2020jensen}, power divergence~\citep{aminian2021information}, and Wasserstein distance~\citep{lopez2018generalization,wang2019information}. An exact characterization of the generalization error for the Gibbs algorithm is provided in \citep{aminian2021exact}. Using rate-distortion theory, \cite{masiha2021learning} and \cite{bu2020information} provide information-theoretic generalization error upper bounds for model misspecification and model compression. Information theoretical approaches are applied mostly to the supervised learning scenario. However, our work offers an information-theoretical upper bound for the generalization error of SSL in the presence of covariate shift.

\section{SSL FRAMEWORK} 
We consider a SSL setting we wish to learn a hypothesis given a set of labeled and unlabeled features. We also wish to use this hypothesis to predict new labels given new features.

We model the features (also known as inputs) using a random variable $X \in \mathcal{X}$ where $\mathcal{X}$ represents the input space; we model the labels (also known as outputs) using a random variable $Y \in \mathcal{Y}$ where $\mathcal{Y}$ represents the output set. We also let $(X^L,Y^L)= \{(X_i^l,Y_i^l)\}_{i=1}^n$ be a training labelled set consisting of a number of input-output data points drawn i.i.d. from $\mathcal{X} \times \mathcal{Y}$ according to $\mu_X^l \otimes P_{Y|X}$, and $X^U = \{X_i^u\}_{i=1}^m$ a training unlabelled set consisting of a number of inputs data point drawn i.i,d. from $\mathcal{X}$ according to the marginal distribution $\mu_X^u$. Note that in the traditional SSL scenario without covariate shift we consider $\mu_X^l=\mu_X^u$.

Under the covariate-shift scenario, we assume that the test and unlabeled feature distribution, $\mu_X^u$, are shifted with respect to labeled inputs distribution, $\mu_X^l$, but the conditional distribution of labels given inputs, $P_{Y|X}$, is the same for test and training dataset.

We represent hypotheses using a random variable $W \in \mathcal{W}$ 
where $\mathcal{W}$ is a hypothesis space. We also represent an SSL algorithm via a Markov kernel that maps a given training set $(Y^L,X^L,X^U)$ onto a hypothesis $W$ of the hypothesis class $\mathcal{W}$ according to the probability law $P_{W|X^L,Y^L,X^U}$.

Let us define the following loss functions:
\begin{itemize}
    \item \textbf{Supervised loss function:} A (non-negative) loss function $\ell:\mathcal{W} \times \mathcal{X} \times \mathcal{Y}  \rightarrow \mathbb{R}^+$ that measures how well a hypothesis predicts a label (output) given a feature (input). 
    \item \textbf{Conditional expectation of supervised loss function:} The expectation of supervised loss function with respect to true conditional distribution, $P_{Y|X}(y|x)$, is defined as follows:
    \begin{equation}\label{Eq: Ideal Unsupervised loss}
        \ell_c(w,x)\triangleq \int_{\mathcal{Y}} \ell(w,x,y)P_{Y|X}(y|x)dy
    \end{equation}
    Note that the conditional distribution, $P_{Y|X}$, is unknown.
    \item \textbf{Unsupervised loss function:} A (non-negative) loss function $\ell_u:\mathcal{X}\times \mathcal{W}\rightarrow \mathbb{R}^+$ that measures the loss related to inputs including unlabeled and labeled features.  
\end{itemize}

We can now define the population risk, the supervised empirical risk, the unsupervised empirical risk and the semi-supervised empirical risk as follows:
\begin{align}\label{Eq: SSL PR}
&L_P(w,P_{X,Y})\triangleq \int_{\mathcal{X}\times \mathcal{Y}}\ell(w,x,y)P_{X,Y}(x,y) dx dy\\\label{Eq: SL ER}
&L_E^{SL}(w,x^L,y^L)\triangleq\frac{1}{n}\sum_{i=1}^n \ell(w,x_i^L,y_i^L)
\\\label{Eq: UL ER}
&L_E^{UL}(w,x^L,x^U)\triangleq\\\nonumber&\quad\frac{1}{n+m}\left(\sum_{i=1}^n \ell_u(w,x_i^L)+ \sum_{j=1}^m \ell_u(w,x_j^U)\right)
\\\label{Eq: SSL ER}
&L_E^{SSL}(w,x^L,y^L,x^U)\triangleq\\\nonumber&\quad\beta L_E^{SL}(w,x^L,y^L) + (1-\beta)L_E^{UL}(W,X^L,X^U),\\\nonumber&\quad  0\le\beta\le1.
\end{align}
Quantify the performance of a hypothesis $w$ delivered by the SSL algorithm on a testing set (population) and the training set, respectively. The hyper-parameter $\beta$ balances between the supervised and unsupervised empirical risk.
\begin{remark}\label{Remark: choice of beta}[Choice of $\beta$]
Choosing $\beta=0$ reduces our problem to an unsupervised learning scenario by considering the unsupervised empirical risk function, $L_E^{UL}(W,X^L,X^U)$, and if we choose $\beta=1$ our problem reduces to a supervised learning setting by considering the supervised empirical risk, $L_E^{SL}(w,x^L,y^L)$.
\end{remark}

\begin{remark}\label{Remark: ideal risk}[Ideal SSL Empirical Risk]
If we substitute the $\ell_u(w,x)$ with $\ell_c(w,x)$ in \eqref{Eq: UL ER}, then the unsupervised empirical risk is an unbiased estimation of population risk~\eqref{Eq: SSL PR} (See appendix~\ref{Proof of remark Ideal SSL ER}). 
\end{remark}
We can also define the generalization error as follows:
\begin{align} \label{Eq:GE}
&\text{gen}(P_{W|X^L,Y^L,X^U},P_{X,Y}) \triangleq\\\nonumber&\quad L_P(w,P_{X,Y})-L_E^{\text{SSL}}(w,x^L,y^L,x^U)
\end{align}
which quantifies how much the population risk deviates from the SSL empirical risk. We can also define the expected generalization error as follows:
\begin{align} \label{eq: expected GE}
&\overline{\text{gen}}(P_{W|X^L,Y^L,X^U},P_{X,Y}) =\\\nonumber &\quad\mathbb{E}_{P_{W,X^L,Y^L,X^U}}[\text{gen}(P_{W|X^L,Y^L,X^U},P_{X,Y})]
\end{align}
For the covariate-shift scenario, we define the generalization error as $\overline{\text{gen}}(P_{W|X^L,Y^L,X^U},\mu_X^u \otimes P_{Y|X}) $ where $\mu_X^u \otimes P_{Y|X}$ is the distribution of test data. 

Now, we will show how our framework reduces to the Pseudo-labeling and entropy minimization. Let us consider a classification task with $q$ labels, i.e., $|\mathcal{Y}|=q$. Suppose that the estimation of true underlying conditional distributions of labels given features and hypothesis, i.e., $\{\widehat{P}_{y=i|w,x}\}_{i=1}^q$, are available. For example, the output of the Softmax layer in a deep neural network could be considered as the estimation of true underlying conditional distributions of labels given features and hypotheses. 

\textbf{Pseudo-labeling:} Consider the log-loss function as supervised loss function and consider the following function
$$ \ell_u(w,x_i)= -\log(\max_j(\widehat{P}_{Y|w,x_i}(y_j|w,x_i)))$$ as unsupervised loss function,
then our framework reduces to the Pseudo-labeling approach in~\citep{lee2013pseudo}.

\textbf{Entropy Minimization:} Let us consider the following unsupervised loss function in classification problem: 
\begin{equation}\label{Eq: classfication real loss}
    \ell_u(w,x_i)=\sum_{j=1}^q \widehat{P}_{Y|w,x_i}(y_j|w,x_i)\ell(w,x_i,y_j).
\end{equation} 
Now, if we choose negative log-loss function as supervised loss function,
$$ \ell(w,x,y)= -\log(\widehat{P}_{Y|w,x}(y|w,x)),$$
then, the unsupervised loss function \eqref{Eq: classfication real loss} would be equal to the conditional entropy,
\begin{align}\label{Eq: entropy min}
    & \ell_u(w,x_i)=H(\widehat{P}_{Y|w,x_i})\\\nonumber
     &=-\sum_{j=1}^q \widehat{P}_{Y|w,x_i}(y_j|w,x_i)\log(\widehat{P}_{Y|w,x_i}(y_j|w,x_i)),
\end{align}
and our framework reduces to the entropy minimization~\citep{grandvalet2005semi}.  

Our framework could be extended by choosing different supervised loss function in \eqref{Eq: classfication real loss}. For example, we could consider the squared log loss~\citep{janocha2016loss}, i.e., $\ell(w,x,y)=-\log^2(\widehat{P}_{Y|w,x}(y|w,x))$, as supervised loss function and unsupervised loss function would be as follows:
\begin{align*}
     &\ell_u(w,x_i)=\\&\quad-\sum_{j=1}^q \widehat{P}_{Y|w,x_i}(y_j|w,x_i)\log^2(\widehat{P}_{Y|w,x_i}(y_j|w,x_i))
\end{align*}
Another classification loss function is $\alpha$-loss~\citep{sypherd2019tunable}, i.e., $\ell(w,x,y)=\frac{\alpha}{\alpha-1}(1-\widehat{P}_{Y|w,x}^{1-1/\alpha}(y|w,x))$ for $\alpha \in (0,\infty)$, and the unsupervised loss function based on $\alpha$-loss would be as follows:
\begin{align*}
     &\ell_u(w,x_i)=\\&\quad\frac{\alpha}{\alpha-1}\sum_{j=1}^q \widehat{P}_{Y|w,x_i}(y_j|w,x_i)(1-\widehat{P}_{Y|w,x_i}^{1-1/\alpha}(y_j|w,x_i)).
\end{align*}

 In entropy minimization~\citep{grandvalet2005semi}, the authors consider solely unlabeled features for conditional entropy. However, in our framework, we also consider the labeled features in the unsupervised empirical risk inspired by Remark~\ref{Remark: ideal risk}. Actually, the labeled features can also help to improve the unsupervised performance of the SSL algorithm. We will show in Section~\ref{sec: experiment}, this helps us to have better performance in comparison to the case considering solely unlabeled features in entropy minimization method.

\section{BOUNDING THE EXPECTED GENERALIZATION ERROR}

We begin by offering an upper bound on the expected generalization error of the SSL scenario under covariate-shift by considering the conditional expectation of supervised loss function instead of unsupervised loss function in \eqref{Eq: SSL ER}. 

\begin{theorem}[Proved in Appendix~\ref{Proof Theorem: SSL upper bound with ideal under covariate-shift}]\label{Theorem: SSL upper bound with ideal under covariate-shift}
Assume that the supervised loss functions, $l(w,x,y)$ is $\sigma_l$-sub-Gaussian~\footnote{A random variable $X$ is $\sigma$-subgaussian if $E[e^{\lambda(X-E[X])}]\leq e^{\frac{\lambda^2 \sigma^2}{2}}$ for all $\lambda \in \mathbb{R}$.} under the $\mu_X^u\otimes P_{Y|X}$ for all $w\in \mathcal{W}$ and $\ell_c(w,x)$ is $\sigma_c$-sub-Gaussian under marginal distribution $\mu_X^u$ for all $w\in \mathcal{W}$. 
The following expected generalization error upper bound under covariate-shift holds:
\begin{align}\label{Eq: upper real F SSL1}
    &|\overline{\text{gen}}(P_{W|X^L,Y^L,X^U},\mu_X^u \otimes P_{Y|X})|\leq\\\nonumber &\beta\sqrt{\frac{2\sigma_l^2}{n}I(W;X^L,Y^L)+2\sigma_l^2 D(\mu_X^l\|\mu_X^u)}
    \\\nonumber&\quad+\frac{n(1-\beta)}{n+m}\sqrt{\frac{2\sigma_c^2}{n}I(W;X^L)+2\sigma_c^2 D(\mu_X^l\|\mu_X^u)}
    \\\nonumber&\quad+\frac{m(1-\beta)}{n+m}\sqrt{\frac{2\sigma_c^2}{m}I(W;X^U)}.
\end{align}
\end{theorem}

If the supervised loss function is bounded in $[a,b]$, then the conditional expectation of supervised loss function, $\ell_c(w,x)$, is also bounded in $[a,b]$ and is $\frac{(b-a)}{2}$-sub-Gaussian under all distributions over $\mathcal{X}$ and all $w\in \mathcal{W}$ and we have $\sigma_l=\sigma_c=\frac{b-a}{2}$.

It is interesting to interpret each term in \eqref{Eq: upper real F SSL1}. The first term,
\begin{equation*}\sqrt{\frac{2\sigma_l^2}{n}I(W;X^L,Y^L)+2\sigma_l^2 D(\mu_X^l\|\mu_X^u)},\end{equation*}
can be interpreted as an upper bound on the supervised learning part of the SSL algorithm. We also have the term $D(\mu_X^l\|\mu_X^u)$, which can be interpreted as the cost of covariate-shift between training and test feature distributions. The second term, \begin{equation*}\frac{n}{n+m}\sqrt{\frac{2\sigma_c^2}{n}I(W;X^L)+2\sigma_c^2 D(\mu_X^l\|\mu_X^u)},\end{equation*} could be interpreted as an upper bound on the unsupervised performance of the SSL algorithm by considering conditional expectation of supervised loss function and labeled features. And finally, $$\frac{m}{n+m}\sqrt{\frac{2\sigma_c^2}{m}I(W;X^U)},$$ could be interpreted as an upper bound on unsupervised performance of the SSL algorithm by considering the unlabeled data.


Now, we provide another expected generalization error upper bound by substituting the conditional expectation of supervised loss function with the unsupervised loss function.
\begin{proposition}[Proved in Appendix~\ref{Proof Prop: SSL upper bound with real under covariate-shift}]\label{Prop: SSL upper bound with real under covariate-shift}
Assume that the supervised loss functions, $l(w,x,y)$ is $\sigma_l$-sub-Gaussian  under the $\mu_X^u\otimes P_{Y|X}$  for all $w\in \mathcal{W}$ and $\ell_u(w,x)$ is $\sigma_u$-sub-Gaussian under marginal distribution $\mu_X^u$ for all $w\in \mathcal{W}$. 
The following upper bound holds on the expected generalization error under covariate-shift by considering the test data distribution as $\mu_X^u \otimes P_{Y|X}$:
\begin{align}\label{Eq: upper bound with real1}
   &|\overline{\text{gen}}(P_{W|X^L,Y^L,X^U},\mu_X^u \otimes P_{Y|X})|\leq\\\nonumber &\beta\sqrt{\frac{2\sigma_l^2}{n}I(W;X^L,Y^L)+2\sigma_l^2D(\mu_X^l\|\mu_X^u)}
    \\\nonumber&\quad+\frac{n(1-\beta)}{n+m}\sqrt{\frac{2\sigma_u^2}{n}I(W;X^L)+2\sigma_u^2 D(\mu_X^l\|\mu_X^u)}
    \\\nonumber&\quad+\frac{m(1-\beta)}{n+m}\sqrt{\frac{2\sigma_u^2}{m}I(W;X^U)}
    \\\nonumber&\quad+ (1-\beta) \Delta^{SSL},
    \\\nn & \text{where}\quad \Delta^{SSL}=\mathbb{E}_{P_{W}\otimes \mu_X^u}[\ell_c(W,X)-\ell_u(W,X)].
\end{align}
\end{proposition}

The term $\Delta^{SSL}$ in \eqref{Eq: upper bound with real1} can be interpreted as the estimation error of conditional distributions of labels given features (prediction uncertainty), \citep{guo2017calibration}, under the learning algorithm. Now we provide an upper bound on the absolute value of $\Delta^{SSL}$ for the classification task.
 \begin{corollary}[Proved in Appendix~\ref{Proof Cor: SSL classification Estimation}]\label{Cor: SSL classification Estimation}
Consider the same assumption as in Proposition~\ref{Prop: SSL upper bound with real under covariate-shift}. We suppose that the supervised loss function $\ell(w,x,y)$ is also $\sigma_u$-sub-Gaussian under distribution $P_{Y|X=x}$ for all $x\in\mathcal{X}$ and $w \in \mathcal{W}$. The following upper bound holds on estimation error of conditional distributions in classification task:
\begin{align}
   &|\Delta^{SSL}|\leq \sqrt{2{\sigma_u}^2 D(\widehat{P}_{Y|W,X}\|P_{Y|X}|P_{W}\otimes \mu_X^u)}.
\end{align}
\end{corollary}
\begin{remark}[Calibration] Based on Corollary~\ref{Cor: SSL classification Estimation}, a poor network calibration would result the looser generalization error upper bound in compare to a calibrated network. The same fact is also discussed by \cite{rizve2020defense}.
\end{remark}

In Proposition~\ref{Prop: SSL upper bound with real under covariate-shift}, if the distribution of $\mu_X^u$ is not absolutely continuous with respect to $\mu_X^l$, then we have $D(\mu_X^l|| \mu_X^u)=\infty$ leading up to a vacuous upper bound. In the following, we therefore propose an alternative upper bound based on total variation that bypasses this issue.

\begin{corollary}[Proved in Appendix~\ref{Proof Cor: total variation}]\label{Cor: total variation}
Assume that the supervised loss functions, $l(w,x,y)$ is bounded in $[0,L_l]$ and $\ell_u(w,x)$ is bounded in $[0,L_u]$. 
The following upper bound holds on the expected generalization error under covariate-shift by considering the test data distribution as $\mu_X^u \otimes P_{Y|X}$:
\begin{align}\label{Eq: upper bound with real}
   &|\overline{\text{gen}}(P_{W|X^L,Y^L,X^U},\mu_X^u \otimes P_{Y|X})|\leq\\\nonumber &\beta\left(\sqrt{\frac{L_l^2}{2n}I(W;X^L,Y^L)}+2L_l  \mathbb{TV}(\mu_X^l,\mu_X^u)\right)
    \\\nonumber&\quad+\frac{n(1-\beta)}{n+m}\left(\sqrt{\frac{L_u^2}{2n}I(W;X^L)}+2L_u \mathbb{TV}(\mu_X^l,\mu_X^u)\right)
    \\\nonumber&\quad+\frac{m(1-\beta)}{n+m}\sqrt{\frac{L_u^2}{2m}I(W;X^U)}
    \\\nonumber&\quad+ (1-\beta) \Delta^{SSL},
\end{align}
where $\Delta^{SSL}=\mathbb{E}_{P_{W}\otimes \mu_X^u}[\ell_c(W,X)-\ell_u(W,X)]$.
\end{corollary}

In Corollary~\ref{Cor: SSL classification Estimation}, if the estimation of conditional probabilities, i.e., $\widehat{P}_{Y|W,X}$, is not absolutely continuous with respect to true conditional probability, i.e., $P_{Y|X}$, for all $x\in\mathcal{X}$ and $w \in \mathcal{W}$, then we have $D(\widehat{P}_{Y|W,X}\|P_{Y|X}|P_{W}\otimes \mu_X^u)=\infty$. In the following Corollary, we derive an upper bound for estimation error of conditional distributions of labels given features, in terms of total variation distance which is bounded.
\begin{corollary}[Proved in Appendix~\ref{Proof cor: est error based on total variation}]\label{cor: est error based on total variation}
Consider the same assumptions as in Corollary~\ref{Cor: total variation},
The following upper bound holds on estimation error of conditional distributions in classification task:
\begin{align}
   &|\Delta^{SSL}|\leq 2L_u \mathbb{TV}(\widehat{P}_{Y|W,X},P_{Y|X}|P_{W}\otimes \mu_X^u),
\end{align}
where $\mathbb{TV}(\widehat{P}_{Y|W,X}\|P_{Y|X}|P_{W}\otimes \mu_X^u)=\mathbb{E}_{P_{W}\otimes \mu_X^u}[\mathbb{TV}(\widehat{P}_{Y|W,X},P_{Y|X})]$.
\end{corollary}
 
It is worthwhile to mention that the results in Theorem~\ref{Theorem: SSL upper bound with ideal under covariate-shift} and Proposition~\ref{Prop: SSL upper bound with real under covariate-shift} could also be applied to the SSL algorithms for traditional SSL scenario (no covariate-shift), where $\mu_X^u=\mu_X^l$.

In the following, we provide a convergence rate for the expected generalization error of SSL algorithms.
\begin{corollary}[Proved in Appendix~\ref{Proof Cor: bounded hypothesis}]\label{Cor: bounded hypothesis}
Consider the same assumptions as in Proposition~\ref{Prop: SSL upper bound with real under covariate-shift} for traditional SSL scenario, $\mu_X^l=\mu_X^u$. Consider also hypothesis space is countable, $|\mathcal{W}|=k$, and $\left|\mathbb{E}_{P_{W}\otimes \mu_X^u}[\ell_c(W,X)-\ell_u(W,X)]\right|\leq \sqrt{2\sigma_u^2}\epsilon$. Then, the following upper bounds holds on the expected generalization error of the SSL algorithm:
\begin{align}\label{Eq: SSL bound hypothesis}
    &\overline{\text{gen}}(P_{W|X^L,Y^L,X^U},P_{X,Y})\leq \beta\sqrt{\frac{2\sigma_l^2 \log(k)}{n}} \\\nonumber &\quad+ (1-\beta)\sqrt{2\sigma_u^2}\left(\sqrt{\frac{ \log(k)}{(n+m)}}+\epsilon\right).
\end{align}
\end{corollary}

 If the estimation error of unsupervised loss function in Corollary~\ref{Cor: bounded hypothesis} is negligible ($\epsilon \rightarrow 0$), the convergence rate of the upper bound in  \eqref{Eq: SSL bound hypothesis}, would be as follows:
 \begin{equation}\label{Eq: convergence rate}
     \mathcal{O}\left(\frac{\beta}{\sqrt{n}}+\frac{(1-\beta)}{\sqrt{n+m}}\right).
 \end{equation}
The convergence rate in \eqref{Eq: convergence rate} depends on the choice of $\beta$. If we consider $\beta=\frac{n}{n+m}$, we have \begin{align}\label{Eq: conv rate}\mathcal{O}\left(\max\left(\frac{\sqrt{n}}{n+m},\frac{m}{(n+m)^{3/2}}\right)\right),\end{align} where shows if $m$ is sufficiently large and $n$ is relatively small, then SSL algorithm's generalization error upper bound's convergence rate would be better than the generalization error upper bound's convergence rate for the supervised learning algorithm, $\mathcal{O}(\frac{1}{\sqrt{n}})$,~\citep{xu2017information}.

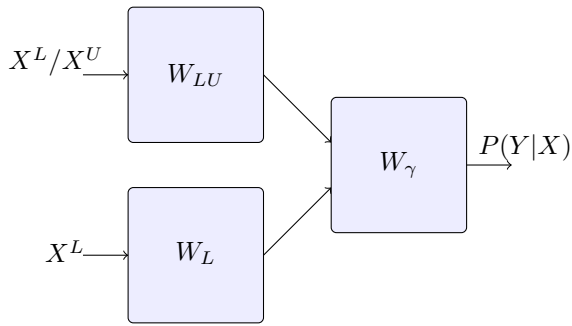
\begin{figure}
\centering
\begin{tikzpicture}[scale=0.6]
\draw [fill=white!93!blue,rounded corners=2pt](0,0) rectangle (3,3);
\node at (1.5,1.55){$W_L$};
\draw[->] (-1,5.5) -- (0,5.5);
\node at (-1.6,5.8){$X^L/X^U$};
\draw[->] (-1,1.5) -- (0,1.5);
\node at (-1.4,1.6){$X^L$};
\draw[->] (3,1.5) -- (4.5,3);
\draw[->] (3,5.5) -- (4.5,4);
\draw [fill=white!93!blue,rounded corners=2pt](0,4) rectangle (3,7);
\node at (1.5,5.5){$W_{LU}$};
\draw [fill=white!93!blue,rounded corners=2pt](4.5,2) rectangle (7.5,5);
\node at (6,3.5){$W_\gamma$};
\draw[->] (7.5,3.5) -- (8.5,3.5);
\node at (8.8,3.9){$P(Y|X)$};
\end{tikzpicture}
\vspace{2mm}
\caption{Structure of The CSSL Method}
\label{fig:sub2}
\end{figure}
\section{CSSL METHOD AND EXPERIMENTS} \label{sec: experiment}
There are two inspirations from our theoretical results. First, as shown in Proposition~\ref{Prop: SSL upper bound with real under covariate-shift} and Corollary~\ref{Cor: SSL classification Estimation}, the estimation of conditional distributions for both labeled and unlabeled data plays an important role in the performance of SSL algorithm. Second, the unsupervised empirical risk $L_E^{UL}$, given in ~\eqref{Eq: UL ER}, is a function of both labeled and unlabeled data and it can help to have the convergence rate as shown in~\eqref{Eq: convergence rate}.

Considering these two inspirations, we now propose the Covariate-shift SSL (CSSL) method (the structure of our model is shown in Figure \ref{fig:sub2}). The unsupervised empirical risk, $L_E^{UL}$, is expressed using the unsupervised loss function \eqref{Eq: entropy min}, which itself is dependent on the conditional distribution estimation. 
And, we have this assumption that the conditional distributions remain invariant for labeled and unlabeled data under covariate-shift. Based on this assumption, in our model, we consider one shared block of parameters (hypothesis), $W_\gamma$, to produce the estimation of conditional probability for both labeled and unlabeled data, i.e., $X^L$ and $X^U$, respectively. As the distribution of $X^L$ and $X^U$ is different, we consider two disjoint blocks of parameters $W_L$ and $W_{LU}$. The input of $W_L$ is only the labeled data, while the inputs to the $W_{LU}$ are both labeled and unlabeled data. If we only use $X^U$ using the loss $L_E^{UL}$ for the training of $W_{LU}$, then the model converges to extreme points (producing only zeros and ones at the output). Hence, feeding $X_L$ to $W_{LU}$ is important to avoid converging to degenerated cases. Based on Figure \ref{fig:sub2}, the  empirical loss defined in \eqref{Eq: SSL ER} can be written as follows, 
\begin{align}
&L_E^{SL}(W_L,W_{LU},W_\gamma,x^L,y^L)=\nonumber\\&\frac{1}{n}\sum_{i=1}^n \ell(W_L,W_\gamma,x_i^L,y_i^L) + \frac{1}{n}\sum_{i=1}^n \ell(W_{LU},W_\gamma,x_i^L,y_i^L)
\\
&L_E^{UL}(W_L,W_{LU},W_{\gamma},x^L,x^U)=\\\nonumber&\quad\frac{1}{n+m}\left(\sum_{i=1}^n \ell_u(W_L,W_{\gamma},x_i^L)+ \sum_{j=1}^m \ell_u(W_{LU},W_{\gamma},x_j^U)\right)
\\
 &L_E^{SSL}(W_L,W_{LU},W_{\gamma},x^L,y^L,x^U)=\\\nonumber &\quad\beta L_E^{SL}(W_L,W_{LU},W_{\gamma},x^L,y^L) +  \\ \nonumber &\quad\quad(1-\beta)L_E^{UL}(W_L,W_{LU},W_{\gamma},x^L,x^U),  0\le\beta\le1
\end{align}

\vspace{-1em}
Now, we show the performance of our CSSL method using two experiments. In the first experiment, we use synthetic data, and in the second experiment, we use in MNIST dataset \citep{MNIST}.
\subsection{Synthetic data}\label{Sec: synthetic data}
In the first experiment, we use the synthetic data generated inspired by the first experiments of \citep{grandvalet2005semi} and \citep{kugelgen2019semi}. We need to create the dataset and impose covariate-shift while satisfying two conditions. First, $p(Y|X)$ should remain constant with the covariate shift. Secondly, we need to make sure that unlabeled data are indeed useful in a semi-supervised learning setup. As discussed in \citep{janzing2015semi} and \citep{kugelgen2019semi}, this can be achieved by ensuring that $(X\rightarrow Y)$ does not hold. This is because, if $(X\rightarrow Y)$ holds then $p(X)$ and $p(Y|X)$ are independent mechanisms \citep{kugelgen2019semi}. Thus, we consider a scenario where we have the following causal learning setting: 
\begin{equation}
    X_C \rightarrow Y \rightarrow X_E.
\end{equation}
Here, $X_C$ denotes the cause features, and $X_E$ denotes the effect features. This scenario frequently arises in practice. For example, in healthcare, $X_C$ can be genetic characteristics, living conditions, etc., $Y$ could represent the illness, and $X_E$ can represent symptoms of the illness like coughing, fever, etc.


In our synthetic dataset, features have dimension of 50. The first 30 features, are the cause features $X_C$, drawn from a mixture of two multivariate Gaussian distributions. Similar to \citep{grandvalet2005semi}, the first Gaussian distribution is $\mathcal N((a_1,\cdots,a_1), s_1 I)$ and the second one is $\mathcal N((-a_1,\cdots,-a_1), s_1 I)$. The mixing probability $\pi$ is $(0.5, 0.5)$. 
The binary label $Y$ is defined as follows
\begin{equation*}
Y = \begin{cases} 1 \quad \text{if}\quad  \epsilon_Y < \sigma(\sum_{i=1}^{30} x_i) \\
0 \quad \text{if}\quad \epsilon_Y > \sigma(\sum_{i=1}^{30} x_i)
\end{cases}, \quad \epsilon_Y \sim U(0,1).
\end{equation*}
Here $\sigma(x)= (1+e^{-x})^{-1}$ is the logistic sigmoid function.
The effect features $X_E$ is of dimension 20, and is defined as 
\be X_E = \begin{cases} a_2 + \epsilon_E \ \quad \text{if}\quad  Y=1 \\
-a_2 + \epsilon_E \quad \text{if}\quad Y=0
\end{cases} \quad \epsilon_Y \sim \mathcal N(0,s_2I).
\ee

\begin{figure}[ht!]
     \centering
     \begin{subfigure}[b]{0.4\textwidth}
         \centering
         \includegraphics[width=\textwidth]{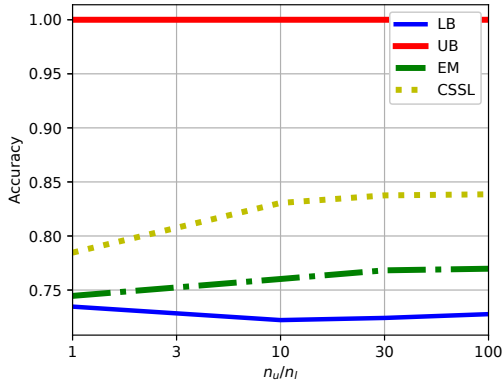}
         \caption{$a_1=0.01$}
         \label{fig:a}
     \end{subfigure}
     \hfill
     \begin{subfigure}[b]{0.4\textwidth}
         \centering
         \includegraphics[width=\textwidth]{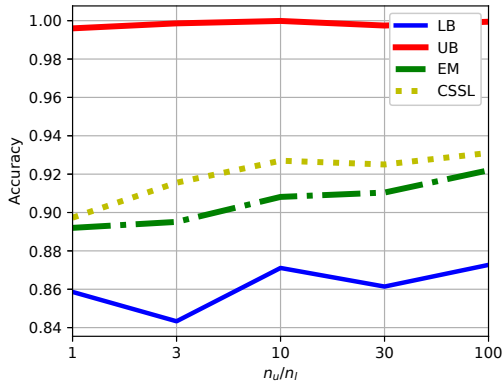}
         \caption{$a_1= 0.03$}
         \label{fig:b}
     \end{subfigure}
        \caption{Comparison of Accuracy of Methods For Two Settings of Synthetic Data}
        \label{fig:exp1}
\end{figure}

Note that variable $a_1$ determines how far apart are the two Gaussian mixtures. As $a_1$ increases the expected value of $\vert\sum_{i=1}^{30} x_i\vert$ increases. This means that $X_C$ will become a better predictor of $Y$. Similarly, $a_2$ determines how good $Y$ can be predicted using $X_E$. The covariate shift will be applied by changing $a_1$. We start with a small value for $a_1$, which means the predictor will rely on $X_E$ for predicting $Y$, and then for the unlabeled features, we increase $a_1$.
Now it is easy to see that both of our conditions are satisfied with this method of data generation. The causal learning setting holds ($X_C \rightarrow Y \rightarrow X_E$) immediately as a consequence of the way we generate data. 
For the other condition we have
\begin{align}
    p(Y|X_C,X_E) &= \frac{p(X_C,X_E|Y)p(Y)}{p(X_C,X_E)} \label{eq:1}\\
    &= \frac{p(X_C|Y)p(X_E|Y)p(Y)}{p(X_C)p(X_E|X_C)}\label{eq:2}\\
    &= \frac{\frac{p(Y|X_C)p(X_C)}{p(Y)}p(X_E|Y)p(Y)}{p(X_C)p(X_E|X_C)}\label{eq:3}\\
    &= \frac{p(Y|X_C)p(X_E|Y)}{\sum_y p(X_E|y,X_C)p(y|X_C)}\label{eq:4}\\
     &= \frac{p(Y|X_C)p(X_E|Y)}{\sum_y p(X_E|y)p(y|X_C)},\label{eq:5}
\end{align}
where \eqref{eq:1} and \eqref{eq:3} hold from Bayes rule, and we used ($X_C \rightarrow Y \rightarrow X_E$) in \eqref{eq:2} and \eqref{eq:5}.
This shows that $p(Y|X_C,X_E)$ remains invariant if we only change $p(X_C)$.

We have used a single layer fully connected network to implement each of $W_L$, $W_{LU}$, and $W_\gamma$. In particular. $W_L$ and $W_{LU}$ are neural networks with an input dimension of 50 and output dimension of 10, and with a ReLU activation function. Whereas $W_\gamma$ gets ten inputs and has two outputs, a softmax function is used at the end to produce the required conditional distributions. The result is reported in Figure \ref{fig:exp1}. The performance of the entropy minimization method \citep{grandvalet2005semi} is also presented. The lower bound is obtained by using only labeled data, and the upper bound is when we used true labels of unlabeled data to train in a supervised manner. We used a three-layer neural network for these two methods, a concatenation of $W_L$ and $W_\gamma$. The value of $\beta$ (and regularization term in EM) can be tuned using ten-fold cross-validation (we have $\beta = 0.02$). In Figure \ref{fig:exp1}, the first figure is corresponding to a scenario where $a_1=0.01$, small $a_1$ means that $X_C$ is not informative and the supervised model relies on $X_E$ for predicting $Y$, we increase $a_1$ significantly for unlabeled data $a_1=0.8$ (thus the upper bound model always predict correctly). In the second figure, we have a more subtle change in $a_1$ and also $X_E$ is more noisy forcing models to consider both $X_E$ and $X_C$ (more details about the experiments is reported in Appendix~\ref{app: experiment details}). In both cases, our proposed method outperforms EM.

\subsection{MNIST}
In this experiment, we use a hand-written digits dataset, MNIST \citep{MNIST}. In order to create covariate shift we impose a selection bias in labeled and unlabeled data. In the labeled data, we choose the majority of images (90 percent of the labeled dataset) from numbers with labels 0 to 4; the remaining 10 percent of the labeled dataset are drawn from images with labels 5 to 9. We reverse this ratio for the unlabeled dataset, with 90 percent of data having labels 5 to 9 and 10 percent with labels 0 to 4. Note that our two conditions in Section \ref{Sec: synthetic data} are satisfied for this experiment. First, it is clear that the conditional distribution of $p(Y|X)$ will not change with the selection bias we imposed. Secondly, no direct causal link exists between $X$ and $Y$. This dataset has been widely used in SSL settings (e.g., in \citep{ganin2016domain}), and it is shown that unlabeled data can indeed improve the performance of the model.

\begin{figure}[t!]
\centering
\includegraphics[scale=0.6]{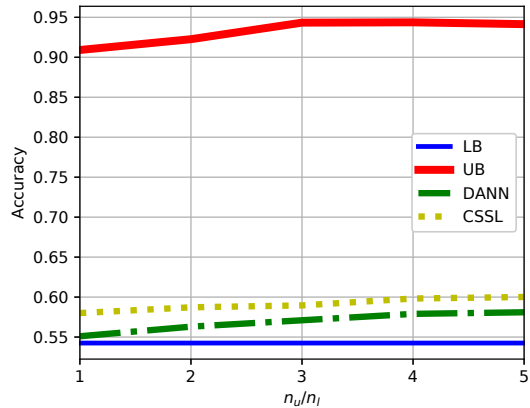} 
\caption{Comparison of Accuracy of CSSL, Domain Adversarial Method (DANN), Lower And Upper Bounds With Varying Ratio of Unlabeled Data For MNIST Dataset.}
\label{fig:exp2}
\end{figure}

In Figure \ref{fig:exp2}, we present the results for this experiment. Similar to the previous experiment, the lower and upper bounds are derived by training a supervised model using only labeled data, and both labeled and unlabeled data (with true labels), respectively. We also report the performance of the domain adversarial method \citep{ganin2016domain} for comparison. Note that in comparison to domain adversarial approach~\citep{ganin2016domain}, the final block $W_\gamma$ in CSSL is trained by both labeled and unlabeled data. However, in domain adversarial approach, the final block is trained solely based on labeled data. Here, we have 1000 labeled images, and we vary the number of unlabeled images (note that because of the limited number of images, we cannot use arbitrarily large numbers of unlabeled data). The networks $W_L$ and $W_{LU}$ have three convolutional layers, and $W_\gamma$ has only one fully connected layer. We used a similar structure for DANN, the feature extractor network is the same as $W_L$ (or $W_{LU}$), and the classifier is similar to $W_\gamma$.

\section{CONCLUSION}
We provide a framework for SSL algorithms that can be reduced to other popular SSL algorithms, including entropy minimization and Pseudo-labeling. Inspired by our framework, we propose new expected generalization error upper bounds based on some information measures distance under the covariate-shift assumption, which illuminates the importance of estimating conditional distributions of labels given features. We also provide an upper bound on the estimation error of conditional distributions. Finally, we propose a method for SSL algorithms under covariate-shift, which outperforms entropy minimization under covariate-shift.
This work motivates further investigation of other supervised loss functions in SSL algorithms. For example, using our framework, we can extend the support vector machine approach based on the Hinge loss function to include unlabeled data. The calibration algorithms~\citep{guo2017calibration} can be applied in our method to see if they will reduce the estimation error of conditional distributions. Our theoretical results and our method are based on covariate-shift assumption, and as a feature work could be extended to other scenarios, e.g., concept drift.

\subsubsection*{Acknowledgements}
We thank the anonymous reviewers for their valuable feedback, which helped us to improve the paper greatly. Gholamali Aminian is supported by the Royal Society Newton International Fellowship, grant no. NIF\textbackslash R1 \textbackslash 192656.
\bibliographystyle{plainnat}
\bibliography{reference}

\begin{thebibliography}{51}
\providecommand{\natexlab}[1]{#1}
\providecommand{\url}[1]{\texttt{#1}}
\expandafter\ifx\csname urlstyle\endcsname\relax
  \providecommand{\doi}[1]{doi: #1}\else
  \providecommand{\doi}{doi: \begingroup \urlstyle{rm}\Url}\fi

\bibitem[Aminian et~al.(2021{\natexlab{a}})Aminian, Bu, Toni, Rodrigues, and
  Wornell]{aminian2021exact}
Gholamali Aminian, Yuheng Bu, Laura Toni, Miguel Rodrigues, and Gregory
  Wornell.
\newblock An exact characterization of the generalization error for the gibbs
  algorithm.
\newblock \emph{Advances in Neural Information Processing Systems}, 34,
  2021{\natexlab{a}}.

\bibitem[Aminian et~al.(2021{\natexlab{b}})Aminian, Toni, and
  Rodrigues]{aminian2020jensen}
Gholamali Aminian, Laura Toni, and Miguel~RD Rodrigues.
\newblock Jensen-shannon information based characterization of the
  generalization error of learning algorithms.
\newblock In \emph{2020 IEEE Information Theory Workshop (ITW)}, pages 1--5.
  IEEE, 2021{\natexlab{b}}.

\bibitem[Aminian et~al.(2021{\natexlab{c}})Aminian, Toni, and
  Rodrigues]{aminian2021information}
Gholamali Aminian, Laura Toni, and Miguel~RD Rodrigues.
\newblock Information-theoretic bounds on the moments of the generalization
  error of learning algorithms.
\newblock In \emph{2021 IEEE International Symposium on Information Theory
  (ISIT)}, pages 682--687. IEEE, 2021{\natexlab{c}}.

\bibitem[Asadi et~al.(2018)Asadi, Abbe, and Verd{\'u}]{asadi2018chaining}
Amir~R Asadi, Emmanuel Abbe, and Sergio Verd{\'u}.
\newblock Chaining mutual information and tightening generalization bounds.
\newblock In \emph{NeurIPS}, 2018.

\bibitem[Ben-David et~al.(2010)Ben-David, Blitzer, Crammer, Kulesza, Pereira,
  and Vaughan]{ben2010theory}
Shai Ben-David, John Blitzer, Koby Crammer, Alex Kulesza, Fernando Pereira, and
  Jennifer~Wortman Vaughan.
\newblock A theory of learning from different domains.
\newblock \emph{Machine learning}, 79\penalty0 (1):\penalty0 151--175, 2010.

\bibitem[Boucheron et~al.(2013)Boucheron, Lugosi, and
  Massart]{boucheron2013concentration}
St{\'e}phane Boucheron, G{\'a}bor Lugosi, and Pascal Massart.
\newblock \emph{Concentration inequalities: A nonasymptotic theory of
  independence}.
\newblock Oxford university press, 2013.

\bibitem[Bu et~al.(2020{\natexlab{a}})Bu, Gao, Zou, and
  Veeravalli]{bu2020information}
Yuheng Bu, Weihao Gao, Shaofeng Zou, and Venugopal Veeravalli.
\newblock Information-theoretic understanding of population risk improvement
  with model compression.
\newblock In \emph{Proceedings of the AAAI Conference on Artificial
  Intelligence}, volume~34, pages 3300--3307, 2020{\natexlab{a}}.

\bibitem[Bu et~al.(2020{\natexlab{b}})Bu, Zou, and
  Veeravalli]{bu2020tightening}
Yuheng Bu, Shaofeng Zou, and Venugopal~V Veeravalli.
\newblock Tightening mutual information based bounds on generalization error.
\newblock \emph{IEEE Journal on Selected Areas in Information Theory},
  2020{\natexlab{b}}.

\bibitem[Chan et~al.(2020)Chan, Alaa, Qian, and Van
  Der~Schaar]{chan2020unlabelled}
Alex Chan, Ahmed Alaa, Zhaozhi Qian, and Mihaela Van Der~Schaar.
\newblock Unlabelled data improves bayesian uncertainty calibration under
  covariate shift.
\newblock In \emph{International Conference on Machine Learning}, pages
  1392--1402. PMLR, 2020.

\bibitem[Chapelle et~al.(2003)Chapelle, Weston, and
  Scholkopf]{chapelle2003cluster}
Olivier Chapelle, Jason Weston, and Bernhard Scholkopf.
\newblock Cluster kernels for semi-supervised learning.
\newblock \emph{Advances in neural information processing systems}, pages
  601--608, 2003.

\bibitem[Chen et~al.(2020)Chen, Wei, Kumar, and Ma]{chen2020self}
Yining Chen, Colin Wei, Ananya Kumar, and Tengyu Ma.
\newblock Self-training avoids using spurious features under domain shift.
\newblock \emph{Advances in Neural Information Processing Systems}, 33, 2020.

\bibitem[Esposito et~al.(2021)Esposito, Gastpar, and
  Issa]{esposito2019generalization}
Amedeo~Roberto Esposito, Michael Gastpar, and Ibrahim Issa.
\newblock Generalization error bounds via r{\'e}nyi-, f-divergences and maximal
  leakage.
\newblock \emph{IEEE Transactions on Information Theory}, 2021.

\bibitem[Ganin et~al.(2016)Ganin, Ustinova, Ajakan, Germain, Larochelle,
  Laviolette, Marchand, and Lempitsky]{ganin2016domain}
Yaroslav Ganin, Evgeniya Ustinova, Hana Ajakan, Pascal Germain, Hugo
  Larochelle, Fran{\c{c}}ois Laviolette, Mario Marchand, and Victor Lempitsky.
\newblock Domain-adversarial training of neural networks.
\newblock \emph{The journal of machine learning research}, 17\penalty0
  (1):\penalty0 2096--2030, 2016.

\bibitem[G{\"o}pfert et~al.(2019)G{\"o}pfert, Ben-David, Bousquet, Gelly,
  Tolstikhin, and Urner]{gopfert2019can}
Christina G{\"o}pfert, Shai Ben-David, Olivier Bousquet, Sylvain Gelly, Ilya
  Tolstikhin, and Ruth Urner.
\newblock When can unlabeled data improve the learning rate?
\newblock In \emph{Conference on Learning Theory}, pages 1500--1518. PMLR,
  2019.

\bibitem[Grandvalet et~al.(2005)Grandvalet, Bengio, et~al.]{grandvalet2005semi}
Yves Grandvalet, Yoshua Bengio, et~al.
\newblock Semi-supervised learning by entropy minimization.
\newblock \emph{CAP}, 367:\penalty0 281--296, 2005.

\bibitem[Guo et~al.(2017)Guo, Pleiss, Sun, and Weinberger]{guo2017calibration}
Chuan Guo, Geoff Pleiss, Yu~Sun, and Kilian~Q Weinberger.
\newblock On calibration of modern neural networks.
\newblock In \emph{International Conference on Machine Learning}, pages
  1321--1330. PMLR, 2017.

\bibitem[Hafez-Kolahi et~al.(2020)Hafez-Kolahi, Golgooni, Kasaei, and
  Soleymani]{hafez2020conditioning}
Hassan Hafez-Kolahi, Zeinab Golgooni, Shohreh Kasaei, and Mahdieh Soleymani.
\newblock Conditioning and processing: Techniques to improve
  information-theoretic generalization bounds.
\newblock \emph{Advances in Neural Information Processing Systems}, 33, 2020.

\bibitem[Haghifam et~al.(2020)Haghifam, Negrea, Khisti, Roy, and
  Dziugaite]{haghifam2020sharpened}
Mahdi Haghifam, Jeffrey Negrea, Ashish Khisti, Daniel~M Roy, and
  Gintare~Karolina Dziugaite.
\newblock Sharpened generalization bounds based on conditional mutual
  information and an application to noisy, iterative algorithms.
\newblock \emph{Advances in Neural Information Processing Systems}, 2020.

\bibitem[He et~al.(2021)He, Yan, and Tan]{he2021information}
Haiyun He, Hanshu Yan, and Vincent~YF Tan.
\newblock Information-theoretic generalization bounds for iterative
  semi-supervised learning.
\newblock \emph{arXiv preprint arXiv:2110.00926}, 2021.

\bibitem[Iscen et~al.(2019)Iscen, Tolias, Avrithis, and Chum]{iscen2019label}
Ahmet Iscen, Giorgos Tolias, Yannis Avrithis, and Ondrej Chum.
\newblock Label propagation for deep semi-supervised learning.
\newblock In \emph{Proceedings of the IEEE/CVF Conference on Computer Vision
  and Pattern Recognition}, pages 5070--5079, 2019.

\bibitem[Janocha and Czarnecki(2016)]{janocha2016loss}
Katarzyna Janocha and Wojciech~Marian Czarnecki.
\newblock On loss functions for deep neural networks in classification.
\newblock \emph{Schedae Informaticae}, 25:\penalty0 49--59, 2016.

\bibitem[Janzing and Sch{\"o}lkopf(2010)]{janzing2010causal}
Dominik Janzing and Bernhard Sch{\"o}lkopf.
\newblock Causal inference using the algorithmic markov condition.
\newblock \emph{IEEE Transactions on Information Theory}, 56\penalty0
  (10):\penalty0 5168--5194, 2010.

\bibitem[Janzing and Sch{\"o}lkopf(2015)]{janzing2015semi}
Dominik Janzing and Bernhard Sch{\"o}lkopf.
\newblock Semi-supervised interpolation in an anticausal learning scenario.
\newblock \emph{The Journal of Machine Learning Research}, 16\penalty0
  (1):\penalty0 1923--1948, 2015.

\bibitem[Kawakita and Kanamori(2013)]{kawakita2013semi}
Masanori Kawakita and Takafumi Kanamori.
\newblock Semi-supervised learning with density-ratio estimation.
\newblock \emph{Machine learning}, 91\penalty0 (2):\penalty0 189--209, 2013.

\bibitem[K{\"u}gelgen et~al.(2019)K{\"u}gelgen, Mey, and
  Loog]{kugelgen2019semi}
Julius K{\"u}gelgen, Alexander Mey, and Marco Loog.
\newblock Semi-generative modelling: Covariate-shift adaptation with cause and
  effect features.
\newblock In \emph{The 22nd International Conference on Artificial Intelligence
  and Statistics}, pages 1361--1369. PMLR, 2019.

\bibitem[LeCun and Cortes(2010)]{MNIST}
Yann LeCun and Corinna Cortes.
\newblock {MNIST} handwritten digit database.
\newblock \emph{public}, 2010.
\newblock URL \url{http://yann.lecun.com/exdb/mnist/}.

\bibitem[Lee et~al.(2013)]{lee2013pseudo}
Dong-Hyun Lee et~al.
\newblock Pseudo-label: The simple and efficient semi-supervised learning
  method for deep neural networks.
\newblock In \emph{Workshop on challenges in representation learning, ICML},
  2013.

\bibitem[Lopez and Jog(2018)]{lopez2018generalization}
Adrian~Tovar Lopez and Varun Jog.
\newblock Generalization error bounds using wasserstein distances.
\newblock In \emph{2018 IEEE Information Theory Workshop (ITW)}, pages 1--5.
  IEEE, 2018.

\bibitem[Mansour et~al.(2009)Mansour, Mohri, and
  Rostamizadeh]{mansour2009domain}
Yishay Mansour, Mehryar Mohri, and Afshin Rostamizadeh.
\newblock Domain adaptation: Learning bounds and algorithms.
\newblock \emph{arXiv preprint arXiv:0902.3430}, 2009.

\bibitem[Masiha et~al.(2021)Masiha, Gohari, Yassaee, and
  Aref]{masiha2021learning}
Mohammad~Saeed Masiha, Amin Gohari, Mohammad~Hossein Yassaee, and Mohammad~Reza
  Aref.
\newblock Learning under distribution mismatch and model misspecification.
\newblock In \emph{IEEE International Symposium on Information Theory (ISIT)},
  2021.

\bibitem[Mey and Loog(2019)]{mey2019improvability}
Alexander Mey and Marco Loog.
\newblock Improvability through semi-supervised learning: A survey of
  theoretical results.
\newblock \emph{arXiv preprint arXiv:1908.09574}, 2019.

\bibitem[Niu et~al.(2013)Niu, Jitkrittum, Dai, Hachiya, and
  Sugiyama]{niu2013squared}
Gang Niu, Wittawat Jitkrittum, Bo~Dai, Hirotaka Hachiya, and Masashi Sugiyama.
\newblock Squared-loss mutual information regularization: A novel
  information-theoretic approach to semi-supervised learning.
\newblock In \emph{International Conference on Machine Learning}, pages 10--18.
  PMLR, 2013.

\bibitem[Oliver et~al.(2018)Oliver, Odena, Raffel, Cubuk, and
  Goodfellow]{oliver2018realistic}
Avital Oliver, Augustus Odena, Colin Raffel, Ekin~D Cubuk, and Ian~J
  Goodfellow.
\newblock Realistic evaluation of deep semi-supervised learning algorithms.
\newblock In \emph{Proceedings of the 32nd International Conference on Neural
  Information Processing Systems}, pages 3239--3250, 2018.

\bibitem[Ouali et~al.(2020)Ouali, Hudelot, and Tami]{ouali2020overview}
Yassine Ouali, C{\'e}line Hudelot, and Myriam Tami.
\newblock An overview of deep semi-supervised learning.
\newblock \emph{arXiv preprint arXiv:2006.05278}, 2020.

\bibitem[Polyanskiy and Wu(2014)]{polyanskiy2014lecture}
Yury Polyanskiy and Yihong Wu.
\newblock Lecture notes on information theory.
\newblock \emph{Lecture Notes for ECE563 (UIUC) and}, 6\penalty0
  (2012-2016):\penalty0 7, 2014.

\bibitem[Rigollet(2007)]{rigollet2007generalization}
Philippe Rigollet.
\newblock Generalization error bounds in semi-supervised classification under
  the cluster assumption.
\newblock \emph{Journal of Machine Learning Research}, 8\penalty0 (7), 2007.

\bibitem[Rizve et~al.(2020)Rizve, Duarte, Rawat, and Shah]{rizve2020defense}
Mamshad~Nayeem Rizve, Kevin Duarte, Yogesh~S Rawat, and Mubarak Shah.
\newblock In defense of pseudo-labeling: An uncertainty-aware pseudo-label
  selection framework for semi-supervised learning.
\newblock In \emph{International Conference on Learning Representations}, 2020.

\bibitem[Russo and Zou(2019)]{russo2019much}
Daniel Russo and James Zou.
\newblock How much does your data exploration overfit? controlling bias via
  information usage.
\newblock \emph{IEEE Transactions on Information Theory}, 66\penalty0
  (1):\penalty0 302--323, 2019.

\bibitem[Ryan and Culp(2015)]{ryan2015semi}
Kenneth~Joseph Ryan and Mark~Vere Culp.
\newblock On semi-supervised linear regression in covariate shift problems.
\newblock \emph{The Journal of Machine Learning Research}, 16\penalty0
  (1):\penalty0 3183--3217, 2015.

\bibitem[Sch{\"o}lkopf et~al.(2012)Sch{\"o}lkopf, Janzing, Peters, Sgouritsa,
  Zhang, and Mooij]{scholkopf2012causal}
Bernhard Sch{\"o}lkopf, Dominik Janzing, Jonas Peters, Eleni Sgouritsa, Kun
  Zhang, and Joris~M Mooij.
\newblock On causal and anticausal learning.
\newblock In \emph{ICML}, 2012.

\bibitem[Shimodaira(2000)]{shimodaira2000improving}
Hidetoshi Shimodaira.
\newblock Improving predictive inference under covariate shift by weighting the
  log-likelihood function.
\newblock \emph{Journal of statistical planning and inference}, 90\penalty0
  (2):\penalty0 227--244, 2000.

\bibitem[Shu et~al.(2018)Shu, Bui, Narui, and Ermon]{shu2018dirt}
Rui Shu, Hung Bui, Hirokazu Narui, and Stefano Ermon.
\newblock A dirt-t approach to unsupervised domain adaptation.
\newblock In \emph{International Conference on Learning Representations}, 2018.

\bibitem[Steinke and Zakynthinou(2020)]{steinke2020reasoning}
Thomas Steinke and Lydia Zakynthinou.
\newblock Reasoning about generalization via conditional mutual information.
\newblock In \emph{Conference on Learning Theory}, pages 3437--3452. PMLR,
  2020.

\bibitem[Sugiyama et~al.(2007)Sugiyama, Krauledat, and
  M{\"u}ller]{sugiyama2007covariate}
Masashi Sugiyama, Matthias Krauledat, and Klaus-Robert M{\"u}ller.
\newblock Covariate shift adaptation by importance weighted cross validation.
\newblock \emph{Journal of Machine Learning Research}, 8\penalty0 (5), 2007.

\bibitem[Sypherd et~al.(2019)Sypherd, Diaz, Cava, Dasarathy, Kairouz, and
  Sankar]{sypherd2019tunable}
Tyler Sypherd, Mario Diaz, John~Kevin Cava, Gautam Dasarathy, Peter Kairouz,
  and Lalitha Sankar.
\newblock A tunable loss function for robust classification: Calibration,
  landscape, and generalization.
\newblock \emph{arXiv preprint arXiv:1906.02314}, 2019.

\bibitem[Wang et~al.(2020)Wang, Shelhamer, Liu, Olshausen, and
  Darrell]{wang2020tent}
Dequan Wang, Evan Shelhamer, Shaoteng Liu, Bruno Olshausen, and Trevor Darrell.
\newblock Tent: Fully test-time adaptation by entropy minimization.
\newblock In \emph{International Conference on Learning Representations}, 2020.

\bibitem[Wang et~al.(2019)Wang, Diaz, Santos~Filho, and
  Calmon]{wang2019information}
Hao Wang, Mario Diaz, Jos{\'e} C{\^a}ndido~S Santos~Filho, and Flavio~P Calmon.
\newblock An information-theoretic view of generalization via wasserstein
  distance.
\newblock In \emph{2019 IEEE International Symposium on Information Theory
  (ISIT)}, pages 577--581. IEEE, 2019.

\bibitem[Wei et~al.(2020)Wei, Shen, Chen, and Ma]{wei2020theoretical}
Colin Wei, Kendrick Shen, Yining Chen, and Tengyu Ma.
\newblock Theoretical analysis of self-training with deep networks on unlabeled
  data.
\newblock In \emph{International Conference on Learning Representations}, 2020.

\bibitem[Xu and Raginsky(2017)]{xu2017information}
Aolin Xu and Maxim Raginsky.
\newblock Information-theoretic analysis of generalization capability of
  learning algorithms.
\newblock In \emph{Advances in Neural Information Processing Systems}, pages
  2524--2533, 2017.

\bibitem[Yang et~al.(2021)Yang, Song, King, and Xu]{yang2021survey}
Xiangli Yang, Zixing Song, Irwin King, and Zenglin Xu.
\newblock A survey on deep semi-supervised learning.
\newblock \emph{arXiv preprint arXiv:2103.00550}, 2021.

\bibitem[Zhu(2020)]{zhu2020semi}
Jingge Zhu.
\newblock Semi-supervised learning: the case when unlabeled data is equally
  useful.
\newblock In \emph{Conference on Uncertainty in Artificial Intelligence}, pages
  709--718. PMLR, 2020.

\end{thebibliography}

\clearpage
\appendix

\thispagestyle{empty}
\onecolumn \makesupplementtitle

\section{SSL Empirical Risk Discussion}\label{Proof of remark Ideal SSL ER}
Let's consider SSL empirical risk based on conditional expectation of supervised loss function as follows:
\begin{align}\label{Eq: SSL ER 1}
L_E(w,x^L,y^L,x^U)\triangleq\frac{\beta}{n}\sum_{i=1}^n \ell(w,x_i^L,y_i^L)
 +\frac{(1-\beta)}{n+m}\left(\sum_{i=1}^n \ell_c(w,x_i^L)+ \sum_{j=1}^m \ell_c(w,x_j^U)\right),  0\le\beta\le1
\end{align}
We have:
\begin{align}\label{Eq: SSL ER 2}
    \mathbb{E}_{P_{XY}}[\ell_c(w,x)]=\mathbb{E}_{P_{XY}}[\ell(w,x,y)]
\end{align}
Using \eqref{Eq: SSL ER 2}, it could be shown that empirical risk based on conditional expectation of supervised loss function \eqref{Eq: SSL ER 1} is an unbiased estimator of population risk:
\begin{align}\label{Eq: SSL ER 3}
\mathbb{E}_{P_{XY}}[L_E(w,x^L,y^L,x^U)]&=\beta \mathbb{E}_{P_{XY}}\left[\frac{1}{n}\sum_{i=1}^n \ell(w,x_i^L,y_i^L)\right]
 +(1-\beta)\mathbb{E}_{P_{XY}}\left[\frac{1}{n+m}\left(\sum_{i=1}^n \ell_c(w,x_i^L)+ \sum_{j=1}^m \ell_c(w,x_j^U)\right)\right]\\\nonumber
 &=\beta \mathbb{E}_{P_{XY}}[\ell(w,x,y)]+(1-\beta)\mathbb{E}_{P_{XY}}[\ell(w,x,y)]\\\nonumber
 & = L_P(w,P_{XY}),
 \quad 0\le\beta\le1
\end{align}
\section{Proof of Theorem~\ref{Theorem: SSL upper bound with ideal under covariate-shift}}\label{Proof Theorem: SSL upper bound with ideal under covariate-shift}

We consider the unsupervised empirical risk functions based on the conditional expectation of supervised loss function, $\ell_c(w,x)$, in the following:

\begin{align}
    &\overline{\text{gen}}(P_{W|X^L,Y^L,X^U},\mu_X^u \otimes P_{Y|X})=\mathbb{E}_{P_{W,X^L,Y^L,X^U}}[L_P(W,\mu_X^u \otimes P_{Y|X})-L_E^{\text{SSL}}(W,X^L,Y^L,X^U)]\\
    &=\mathbb{E}_{P_{W,X^L,Y^L,X^U}}[L_P(W,\mu_X^u \otimes P_{Y|X})-\beta L_E^{SL}(W,X^L,Y^L) - (1-\beta)L_E^{UL}(W,X^L,X^U)]=\\
    &\beta(\mathbb{E}_{P_{W,X^L,Y^L}}[L_P(W,\mu_X^u \otimes P_{Y|X})-L_E^{SL}(W,X^L,Y^L)]\\\nn & +(1-\beta)\mathbb{E}_{P_{W,X^U,X^L}}[L_P(W,\mu_X^u \otimes P_{Y|X})-L_E^{UL}(W,X^L,X^U)]=\\
    &\beta\mathbb{E}_{P_{W,X^L,Y^L}}[L_P(W,\mu_X^u \otimes P_{Y|X})-L_E^{SL}(W,X^L,Y^L)]\\\nn &
    +(1-\beta)\frac{n}{m+n}\mathbb{E}_{P_{W,X^L}}[L_P(W,\mu_X^u \otimes P_{Y|X})-L_E^{UL}(W,X^L)]\\\nn &
    +(1-\beta)\frac{m}{m+n}\mathbb{E}_{P_{W,X^U}}[L_P(W,\mu_X^u \otimes P_{Y|X})-L_E^{UL}(W,X^U)]\\
    &=\beta\left(\mathbb{E}_{P_{W}\otimes (\mu_X^u \otimes P_{Y|X})^{\otimes n}}[L_E^{SL}(W,X^L,Y^L)]-\mathbb{E}_{P_{W,X^L,Y^L}}[L_E^{SL}(W,X^L,Y^L)]\right)\\\nn &
    +(1-\beta)\frac{n}{m+n}\left(\mathbb{E}_{P_{W}\otimes {\mu_X^u}^{ \otimes n}}[L_E^{UL}(W,X^L)]-\mathbb{E}_{P_{W,X^L}}[L_E^{UL}(W,X^L)]\right)\\\nn &
    +(1-\beta)\frac{m}{m+n}\left(\mathbb{E}_{P_{W}\otimes {\mu_X^u}^{ \otimes m}}[L_E^{UL}(W,X^U)]-\mathbb{E}_{P_{W,X^U}}[L_E^{UL}(W,X^U)]\right),\displaybreak
    \end{align}
    where $L_E^{UL}(W,X^L)=\frac{1}{n}\sum_{i=1}^n \ell_c(w,x_i^L)$, $L_E^{UL}(W,X^U)=\frac{1}{m}\sum_{j=1}^m \ell_c(w,x_j^U)$, $P_{X^L}={\mu_X^l}^{\otimes n}$ and $P_{X^U}={\mu_X^u}^{ \otimes m}$. Now we have:
    
    \begin{align}
    &\beta\left(\mathbb{E}_{P_{W}\otimes (\mu_X^u \otimes P_{Y|X})^{\otimes n}}[L_E^{SL}(W,X^L,Y^L)]-\mathbb{E}_{P_{W,X^L,Y^L}}[L_E^{SL}(W,X^L,Y^L)]\right)\\\nn &
    +(1-\beta)\frac{n}{m+n}\left(\mathbb{E}_{P_{W}\otimes {\mu_X^u}^{ \otimes n}}[L_E^{UL}(W,X^L)]-\mathbb{E}_{P_{W,X^L}}[L_E^{UL}(W,X^L)]\right)\\\nn &
    +(1-\beta)\frac{m}{m+n}\left(\mathbb{E}_{P_{W}\otimes {\mu_X^u}^{ \otimes m}}[L_E^{UL}(W,X^U)]-\mathbb{E}_{P_{W,X^U}}[L_E^{UL}(W,X^U)]\right)\\
    &\leq \beta\left|\mathbb{E}_{P_{W,X^L,Y^L}}[L_E^{SL}(W,X^L,Y^L)]-\mathbb{E}_{P_{W}\otimes (\mu_X^u \otimes P_{Y|X})^{\otimes n}}[L_E^{SL}(W,X^L,Y^L)]\right|\\\nn &
    +(1-\beta)\frac{n}{m+n}\left|\mathbb{E}_{P_{W,X^L}}[L_E^{UL}(W,X^L)]-\mathbb{E}_{P_{W}\otimes {\mu_X^u}^{ \otimes n}}[L_E^{UL}(W,X^L)]\right|\\\nn &
    +(1-\beta)\frac{m}{m+n}\left|\mathbb{E}_{P_{W,X^U}}[L_E^{UL}(W,X^U)]-\mathbb{E}_{P_{W}\otimes {\mu_X^u}^{ \otimes m}}[L_E^{UL}(W,X^U)]\right|\\\label{Eq: based on donsker1}
    &\leq \beta\sqrt{\frac{2 \sigma_l^2 D(P_{W,X^L,Y^L} \| P_{W}\otimes (\mu_X^u \otimes P_{Y|X})^{\otimes n})}{n}}\\\nn &
    +(1-\beta)\frac{n}{m+n}\sqrt{\frac{2 \sigma_c^2 D(P_{W,X^L} \| P_{W}\otimes {\mu_X^u}^{ \otimes n})}{n}}\\\nn &
    +(1-\beta)\frac{m}{m+n}\sqrt{\frac{2 \sigma_c^2 D(P_{W,X^U} \| P_{W}\otimes {\mu_X^u}^{ \otimes m})}{m}}
    \\\label{Eq: KL decompose1}
    &\leq\beta\sqrt{\frac{2\sigma_l^2}{n}I(W;X^L,Y^L)+2\sigma_l^2 D(\mu_X^l\|\mu_X^u)}\\\nn &
    +\frac{n(1-\beta)}{n+m}\sqrt{\frac{2\sigma_c^2}{n}I(W;X^L)+2\sigma_c^2 D(\mu_X^l\|\mu_X^u)}\\\nn &
    +\frac{m(1-\beta)}{n+m}\sqrt{\frac{2\sigma_c^2}{m}I(W;X^U)}
\end{align}

 The result~\eqref{Eq: based on donsker1} follows from Donsker-Varadhan representation of KL divergence~\citep{xu2017information} and the result~\eqref{Eq: KL decompose1} follows from the fact that
\begin{equation}
    D(P_{W,X^L}\|{\mu_X^u}^{\otimes n} \otimes P_W)=D(P_{X^L,W}\|{\mu_X^l}^{\otimes n} \otimes P_W)+ n D(\mu_X^l\|\mu_X^u)=I(W;X^L)+n D(\mu_X^l\|\mu_X^u)
\end{equation}

\section{Proof of Proposition~\ref{Prop: SSL upper bound with real under covariate-shift}}\label{Proof Prop: SSL upper bound with real under covariate-shift}

We consider the unsupervised empirical risk functions based on the unsupervised loss function in the following.
\begin{align}
    &\overline{\text{gen}}(P_{W|X^L,Y^L,X^U},\mu_X^u \otimes P_{Y|X})=L_P(W,\mu_X^u \otimes P_{Y|X})-L_E^{\text{SSL}}(W,X^L,Y^L,X^U)\\
    &=\mathbb{E}_{P_{W,X^L,Y^L,X^U}}[L_P(W,\mu_X^u \otimes P_{Y|X})-\beta L_E^{SL}(W,X^L,Y^L) - (1-\beta)L_E^{UL}(W,X^L,X^U)]
    \\&
    =\beta\mathbb{E}_{P_{W,X^L,Y^L}}[L_P(W,\mu_X^u \otimes P_{Y|X})-L_E^{SL}(W,X^L,Y^L)]\\\nn & +(1-\beta)\mathbb{E}_{P_{W,X^U,X^L}}[L_P(W,\mu_X^u \otimes P_{Y|X})-L_E^{UL}(W,X^L,X^U)]\\
    &=\beta\mathbb{E}_{P_{W,X^L,Y^L}}[L_P(W,\mu_X^u \otimes P_{Y|X})-L_E^{SL}(W,X^L,Y^L)]\\\nn &
    +(1-\beta)\frac{n}{m+n}\mathbb{E}_{P_{W,X^L}}[L_P(W,\mu_X^u \otimes P_{Y|X})-L_E^{UL}(W,X^L)]\\\nn &
    +(1-\beta)\frac{m}{m+n}\mathbb{E}_{P_{W,X^U}}[L_P(W,\mu_X^u \otimes P_{Y|X})-L_E^{UL}(W,X^U)]
    \\
    &=\beta\left(\mathbb{E}_{P_{W}\otimes (\mu_X^u \otimes P_{Y|X})^{\otimes n}}[L_E^{SL}(W,X^L,Y^L)]-\mathbb{E}_{P_{W,X^L,Y^L}}[L_E^{SL}(W,X^L,Y^L)]\right)\\\nn &
    +(1-\beta)\frac{n}{m+n}\left(\mathbb{E}_{P_{W}\otimes {\mu_X^u}^{ \otimes n}}[L_E^{UL}(W,X^L)]-\mathbb{E}_{P_{W,X^L}}[L_E^{UL}(W,X^L)]\right)\\\nn &
    +(1-\beta)\frac{m}{m+n}\left(\mathbb{E}_{P_{W}\otimes {\mu_X^u}^{ \otimes m}}[L_E^{UL}(W,X^U)]-\mathbb{E}_{P_{W,X^U}}[L_E^{UL}(W,X^U)]\right)\\\nn &
    +(1-\beta)(\mathbb{E}_{P_W \otimes \mu_X^u}[\ell_c(W,X)-\ell_u(W,X)])\\
    &\leq \beta\left|\mathbb{E}_{P_{W,X^L,Y^L}}[L_E^{SL}(W,X^L,Y^L)]-\mathbb{E}_{P_{W}\otimes (\mu_X^u \otimes P_{Y|X})^{\otimes n}}[L_E^{SL}(W,X^L,Y^L)]\right|\\\nn &
    +(1-\beta)\frac{n}{m+n}\left|\mathbb{E}_{P_{W,X^L}}[L_E^{UL}(W,X^L)]-\mathbb{E}_{P_{W}\otimes {\mu_X^u}^{ \otimes n}}[L_E^{UL}(W,X^L)]\right|\\\nn &
    +(1-\beta)\frac{m}{m+n}\left|\mathbb{E}_{P_{W,X^U}}[L_E^{UL}(W,X^U)]-\mathbb{E}_{P_{W}\otimes {\mu_X^u}^{ \otimes m}}[L_E^{UL}(W,X^U)]\right|\\\nn &
    +(1-\beta)\left|\mathbb{E}_{P_W \otimes \mu_X^u}[\ell_c(W,X)-\ell_u(W,X)]\right|\\\label{Eq: based on donsker2}
    &\leq \beta\sqrt{\frac{2 \sigma_l^2 D(P_{W,X^L,Y^L} \| P_{W}\otimes (\mu_X^u \otimes P_{Y|X})^{\otimes n})}{n}}\\\nn &
    +(1-\beta)\frac{n}{m+n}\sqrt{\frac{2 \sigma_u^2 D(P_{W,X^L} \| P_{W}\otimes {\mu_X^u}^{ \otimes n})}{n}}\\\nn &
    +(1-\beta)\frac{m}{m+n}\sqrt{\frac{2 \sigma_u^2 D(P_{W,X^U} \| P_{W}\otimes {\mu_X^u}^{ \otimes m})}{m}}\\\nn &
    +(1-\beta)\left|\mathbb{E}_{P_W \otimes \mu_X^u}[\ell_c(W,X)-\ell_u(W,X)]\right|
    \\\label{Eq: KL decompose2}
    &\leq\beta\sqrt{\frac{2\sigma_l^2}{n}I(W;X^L,Y^L)+2\sigma_l^2 D(\mu_X^l\|\mu_X^u)}\\\nn &
    +\frac{n(1-\beta)}{n+m}\sqrt{\frac{2\sigma_u^2}{n}I(W;X^L)+2\sigma_c^2 D(\mu_X^l\|\mu_X^u)}\\\nn &
    +\frac{m(1-\beta)}{n+m}\sqrt{\frac{2\sigma_u^2}{m}I(W;X^U)}\\\nn &
    +(1-\beta)\left|\mathbb{E}_{P_W \otimes \mu_X^u}[\ell_c(W,X)-\ell_u(W,X)]\right|,
\end{align}
where $L_E^{UL}(W,X^L)=\frac{1}{n}\sum_{i=1}^n \ell_u(w,x_i^L)$, $L_E^{UL}(W,X^U)=\frac{1}{m}\sum_{j=1}^m \ell_u(w,x_j^U)$, $P_{X^L}={\mu_X^l}^{\otimes n}$ and $P_{X^U}={\mu_X^u}^{ \otimes m}$. \eqref{Eq: based on donsker2} follows from Donsker-Varadhan representation of KL divergence. \eqref{Eq: KL decompose2} follows from the fact that
\begin{equation}
    D(P_{X,W}\|\mu_X^u \otimes P_W)=D(P_{X,W}\|P_X \otimes P_W)+ D(\mu_X^l\|\mu_X^u)=I(W;X)+D(\mu_X^l\|\mu_X^u)
\end{equation}

\section{Proof of Corollary~\ref{Cor: SSL classification Estimation}}\label{Proof Cor: SSL classification Estimation}
For the classification task~\eqref{Eq: classfication real loss}, we have:
\begin{align}
    &\left|\Delta^{SSL}\right|=\left|\mathbb{E}_{P_W\otimes \mu_X^u}[\ell_c(W,X)-\ell_u(W,X)]\right|=\\
    &\left|\mathbb{E}_{P_W\otimes \mu_X^u}\left[ \sum_{j=1}^q \widehat{P}_{Y=y_j|W,X}\ell(W,X,y_j)-\sum_{j=1}^q P_{Y=y_j|X}\ell(W,X,y_j)\right]\right|\\\label{Eq: donsker for calibration}
    &\leq \sqrt{2\sigma_u^2 D(P_W\otimes \mu_X^u\otimes P_{Y|X}\|P_W\otimes \mu_X^u\otimes \widehat{P}_{Y|W,X})}\\
    &=\sqrt{2\sigma_u^2 D( P_{Y|X}\| \widehat{P}_{Y|W,X}|P_W\otimes \mu_X^u)},
\end{align}
where \eqref{Eq: donsker for calibration} is based on Donsker-Varadhan representation of KL divergence, \citep{boucheron2013concentration}.

\section{Proof of Corollary~\ref{Cor: total variation}}\label{Proof Cor: total variation}

We consider the unsupervised empirical risk functions based on the unsupervised loss function in the following.
\begin{align}
    &\overline{\text{gen}}(P_{W|X^L,Y^L,X^U},\mu_X^u \otimes P_{Y|X})=L_P(W,\mu_X^u \otimes P_{Y|X})-L_E^{\text{SSL}}(W,X^L,Y^L,X^U)
    \\
    &=\mathbb{E}_{P_{W,X^L,Y^L,X^U}}[L_P(W,\mu_X^u \otimes P_{Y|X})-\beta L_E^{SL}(W,X^L,Y^L) - (1-\beta)L_E^{UL}(W,X^L,X^U)]
    \\
    &=\beta\mathbb{E}_{P_{W,X^L,Y^L}}[L_P(W,\mu_X^u \otimes P_{Y|X})-L_E^{SL}(W,X^L,Y^L)]
    \\\nn & +(1-\beta)\mathbb{E}_{P_{W,X^U,X^L}}[L_P(W,\mu_X^u \otimes P_{Y|X})-L_E^{UL}(W,X^L,X^U)]
    \\
    &=\beta\mathbb{E}_{P_{W,X^L,Y^L}}[L_P(W,\mu_X^u \otimes P_{Y|X})-L_E^{SL}(W,X^L,Y^L)]
    \\\nn &
    +(1-\beta)\frac{n}{m+n}\mathbb{E}_{P_{W,X^L}}[L_P(W,\mu_X^u \otimes P_{Y|X})-L_E^{UL}(W,X^L)]
    \\\nn &
    +(1-\beta)\frac{m}{m+n}\mathbb{E}_{P_{W,X^U}}[L_P(W,\mu_X^u \otimes P_{Y|X})-L_E^{UL}(W,X^U)]
    \\
    &=\beta\left(\mathbb{E}_{P_{W}\otimes (\mu_X^u \otimes P_{Y|X})^{\otimes n}}[L_E^{SL}(W,X^L,Y^L)]-\mathbb{E}_{P_{W,X^L,Y^L}}[L_E^{SL}(W,X^L,Y^L)]\right)
    \\\nn &
    +(1-\beta)\frac{n}{m+n}\left(\mathbb{E}_{P_{W}\otimes {\mu_X^u}^{ \otimes n}}[L_E^{UL}(W,X^L)]-\mathbb{E}_{P_{W,X^L}}[L_E^{UL}(W,X^L)]\right)
    \\\nn &
    +(1-\beta)\frac{m}{m+n}\left(\mathbb{E}_{P_{W}\otimes {\mu_X^u}^{ \otimes m}}[L_E^{UL}(W,X^U)]-\mathbb{E}_{P_{W,X^U}}[L_E^{UL}(W,X^U)]\right)
    \\\nn &
    +(1-\beta)(\mathbb{E}_{P_W \otimes \mu_X^u}[\ell_c(W,X)-\ell_u(W,X)])
    \\
    &\leq \beta\left|\mathbb{E}_{P_{W,X^L,Y^L}}[L_E^{SL}(W,X^L,Y^L)]-\mathbb{E}_{P_{W}\otimes (\mu_X^l \otimes P_{Y|X})^{\otimes n}}[L_E^{SL}(W,X^L,Y^L)]\right|
    \\\nn &
    +\beta\left|\mathbb{E}_{P_{W}\otimes (\mu_X^l \otimes P_{Y|X})^{\otimes n}}[L_E^{SL}(W,X^L,Y^L)]- \mathbb{E}_{P_{W}\otimes (\mu_X^u \otimes P_{Y|X})^{\otimes n}}[L_E^{SL}(W,X^L,Y^L)]\right|
    \\\nn &
    +(1-\beta)\frac{n}{m+n}\left|\mathbb{E}_{P_{W,X^L}}[L_E^{UL}(W,X^L)]-\mathbb{E}_{P_{W}\otimes {\mu_X^u}^{ \otimes n}}[L_E^{UL}(W,X^L)]\right|
    \\\nn &
    +(1-\beta)\frac{n}{m+n}\left|\mathbb{E}_{P_{W}\otimes {\mu_X^l}^{ \otimes n}}[L_E^{UL}(W,X^L)]-\mathbb{E}_{P_{W}\otimes {\mu_X^u}^{ \otimes n}}[L_E^{UL}(W,X^L)]\right|
    \\\nn &
    +(1-\beta)\frac{m}{m+n}\left|\mathbb{E}_{P_{W,X^U}}[L_E^{UL}(W,X^U)]-\mathbb{E}_{P_{W}\otimes {\mu_X^u}^{ \otimes m}}[L_E^{UL}(W,X^U)]\right|
    \\\nn &
    +(1-\beta)\left|\mathbb{E}_{P_W \otimes \mu_X^u}[\ell_c(W,X)-\ell_u(W,X)]\right|\\\label{Eq: based on donsker3}
    &\leq \beta\left(\sqrt{\frac{ L_l^2 I(W;X^L,Y^L)}{2n}}+2L_l \mathbb{TV}(\mu_X^l,\mu_X^u)\right)
    \\\nn &
    +\frac{n(1-\beta)}{m+n}\left(\sqrt{\frac{ L_u^2 I(W;X^L)}{2n}}+2L_u \mathbb{TV}(\mu_X^l,\mu_X^u)\right)
    \\\nn &
    +(1-\beta)\frac{m}{m+n}\sqrt{\frac{ L_u^2 I(W;X^U)}{2m}}
    \\\nn &
    +(1-\beta)\left|\mathbb{E}_{P_W \otimes \mu_X^u}[\ell_c(W,X)-\ell_u(W,X)]\right|,
\end{align}
where $L_E^{UL}(W,X^L)=\frac{1}{n}\sum_{i=1}^n \ell_u(w,x_i^L)$, $L_E^{UL}(W,X^U)=\frac{1}{m}\sum_{j=1}^m \ell_u(w,x_j^U)$, $P_{X^L}={\mu_X^l}^{\otimes n}$ and $P_{X^U}={\mu_X^u}^{ \otimes m}$. \eqref{Eq: based on donsker3} follows from Donsker-Varadhan representation of KL divergence, \citep{boucheron2013concentration}, and the variational representation of total variation~\eqref{Eq: tv rep}, 

\begin{align}
    &\left|\mathbb{E}_{P_{W}\otimes (\mu_X^l \otimes P_{Y|X})^{\otimes n}}[L_E^{SL}(W,X^L,Y^L)]- \mathbb{E}_{P_{W}\otimes (\mu_X^u \otimes P_{Y|X})^{\otimes n}}[L_E^{SL}(W,X^L,Y^L)]\right|
    \\\nn &
    \leq \frac{1}{n}\sum_{i=1}^n|\mathbb{E}_{P_{W}\otimes \mu_X^l \otimes P_{Y_i|X_i}}[\ell(W,X_i,Y_i)]- \mathbb{E}_{P_{W}\otimes \mu_X^u \otimes P_{Y_i|X_i}}[\ell(W,X_i,Y_i)]| \\\nn &
    \leq \frac{1}{n}\sum_{i=1}^n \mathbb{TV}(P_{W}\otimes \mu_X^l \otimes P_{Y_i|X_i},P_{W}\otimes \mu_X^u \otimes P_{Y_i|X_i})\\\nn &
    =\mathbb{TV}(\mu_X^l,\mu_X^u).
\end{align}

\section{Proof of Corollary~\ref{cor: est error based on total variation}}\label{Proof cor: est error based on total variation}

For the classification task~\eqref{Eq: classfication real loss}, we have:
\begin{align}
    &\left|\Delta^{SSL}\right|=\left|\mathbb{E}_{P_W\otimes \mu_X^u}[\ell_c(W,X)-\ell_u(W,X)]\right|\\
    &=\left|\mathbb{E}_{P_W\otimes \mu_X^u}\left[ \sum_{j=1}^q \widehat{P}_{Y=y_j|W,X}\ell(W,X,y_j)-\sum_{j=1}^q P_{Y=y_j|X}\ell(W,X,y_j)\right]\right|\\\label{Eq: Total variation}
    &\leq  2 L_u \mathbb{E}_{P_W\otimes \mu_X^u}\left[ \mathbb{TV}(\widehat{P}_{Y|W,X},P_{Y|X})\right],
\end{align}
where \eqref{Eq: donsker for calibration} is based on the variational representation of total variation~\eqref{Eq: tv rep}.

\section{Proof of Corollary~\ref{Cor: bounded hypothesis}}\label{Proof Cor: bounded hypothesis}
Considering $H(W)\le \log(k)$ and the fact that the $H(W)$ is the upper bound on the mutual information between $W$ and any other random variable, we have:
\begin{align}
   &|\overline{\text{gen}}(P_{W|X^L,Y^L,X^U},\mu_X^u \otimes P_{Y|X})|\\\nonumber &\leq\beta\sqrt{\frac{2\sigma_l^2}{n}I(W;X^L,Y^L)}
   +\frac{n(1-\beta)}{n+m}\sqrt{\frac{2\sigma_u^2}{n}I(W;X^L)}
    +\frac{m(1-\beta)}{n+m}\sqrt{\frac{2\sigma_u^2}{m}I(W;X^U)}
   + (1-\beta) \Delta^{SSL}\\
   &\le \beta\sqrt{\frac{2\sigma_l^2}{n}I(W;X^L,Y^L)}
   +(1-\beta)\sqrt{\frac{2\sigma_u^2}{n+m}I(W;X^L,X^U)}
   + (1-\beta) \sqrt{2\sigma_u^2}\epsilon\\
   &\le \beta\sqrt{\frac{2\sigma_l^2}{n}H(W)}
   +(1-\beta)\sqrt{\frac{2\sigma_u^2}{n+m}H(W)}
   + (1-\beta) \sqrt{2\sigma_u^2}\epsilon\\
   &\le \beta\sqrt{\frac{2\sigma_l^2}{n}\log(k)}
   +(1-\beta)\sqrt{2\sigma_u^2}\left(\sqrt{\frac{\log(k)}{n+m}}+\epsilon\right).
\end{align}


\section{Experiment Details}\label{app: experiment details}
We have used pytorch package for implementation of the code. For the first experiment (synthetic data), a single layer fully connected network is used to implement each of $W_L$, $W_{LU}$, and $W_\gamma$ networks. In particular. $W_L$ and $W_{LU}$ are neural networks with an input dimension of 50 and output dimension of 10, and with a ReLU activation function. $W_\gamma$ has ten inputs and two output nodes, a softmax function is used at the end to produce the required conditional distributions. The value of $\beta$ (and regularization term in EM) can be tuned using ten-fold cross-validation, we have $\beta = 0.02$. Number of labeled data points for both figures is 300.

For Figure \ref{fig:a}, we have used the following parameters for producing the dataset: 
$a_1=0.01$, $s_1=0.05$, $a_2=0.01$, and $s_2 = 2$. For the unlabeled data and test data we chose $a_1=0.8$. Since, $a_1$ is large for unlabeled data, $X_C$ will become a very good predictor of $Y$, thus the upper bound model always predict the output correctly.

For Figure \ref{fig:b}, we used these parameters for synthetic data: 
$a_1=0.03$, $s_1=0.05$, $a_2=0.01$, and $s_2 = 0.2$. Note, that here $X_C$ is more informative in labeled data. Also, the variance of $X_E$ is decreased, hence it becomes more useful and hence the lower bound has improved. 
We have made a more subtle change in $a_1$ for unlabeled data, we chose $a_1 = 0.3$ for unlabeled and test data. As a result upper bound is not always 1 anymore.
`
\subsection{Different distribution for unlabeled and test data}
Here, we evaluate the performance of different methods when the distribution of test data is different from unlabeled data. When the data is collected sequentially it is possible that the distribution constantly changes, hence it is possible that the distribution of test data does not match with unlabeled data. In fact, it is an interesting direction for future work to consider non-stationary data stream, where the distribution of data constantly changes.

In Table \ref{exp3-table}, we consider the setup of first experiment (Figure \ref{fig:a}), while changing $a_1$ for the test data (recall that $a_1=0.8$ for unlabeled data). We repeat the experiment five times and report the mean and standard deviation of the accuracy of each method. We have 300 labeled data and 3000 unlabeled data. It can be seen that CSSL outperforms EM.
\clearpage
\begin{table}[ht!]
\centering
\caption{Comparison of different methods when the distribution of unlabeled and test data is different}
{\scriptsize
\begin{tabular}{cccc} 
\toprule
Value of $a_1$ for test dataset& $a_1=0.9$ & $a_1=0.6$ & $a_1=0.4$ \\ \midrule
Lower bound & $  0.761\pm  0.011 $ &$ 0.681\pm 0.010  $ & $ 0.621\pm 0.010  $\\\midrule
EM & $   0.874 \pm  0.032 $& $  0.781 \pm 0.036 $ &$ 0.705 \pm 0.030 $ \\\midrule
CSSL & $  \bm{0.876\pm  0.070}$ & $ \bm{0.792\pm 0.065}$&$ \bm{ 0.712\pm  0.053 }$ \\\midrule
Upper bound &$  0.999\pm 0.0001 $ & $ 0.997\pm0.004 $&$ 0.977\pm 0.024 $ \\
\bottomrule 
\end{tabular}}
\label{exp3-table}
\vspace{-4pt}
\end{table}

\end{document}